\documentstyle[aps,prl, epsf]{revtex}

\newcommand{\eqref}[1]{Eq.~(\protect\ref{#1})}
\newcommand{\figref}[1]{Fig.~\protect\ref{#1}}

\begin{document}

\draft

\wideabs{
\title{
Dynamics and Selection of Giant Spirals in
Rayleigh-B\'{e}nard Convection
}

\author{
Brendan~B. Plapp\cite{A-address}, David~A. Egolf\cite{L-address},
and Eberhard Bodenschatz\cite{eb-email}
}

\address{
Laboratory of Atomic and Solid State Physics, Cornell University,
Ithaca, NY 14853-2501
}

\author{
Werner Pesch
}

\address{
Physikalisches Institut der Universit\"{a}t Bayreuth,
D-95440 Bayreuth, Germany}

\date{\today}

\maketitle

\begin{abstract}

For Rayleigh-B\'{e}nard convection of a fluid with Prandtl number
$\sigma \approx 1$, we report experimental and theoretical results on a
pattern selection mechanism for cell-filling, giant,  rotating spirals.
We show that the pattern selection in a certain limit can be explained
quantitatively by a phase-diffusion mechanism.  This mechanism for
pattern selection is very
different from that for spirals in excitable media.
\end{abstract}

\pacs{
47.54.+r,
47.52.+j,
47.20.Lz,
47.27.Te
}

}

\narrowtext

Spiral patterns are found in many pattern-forming
systems\cite{CrossHohenberg93}. Famous examples include cardiac
arrhythmias of the heart \cite{PhysicsTodayAug96}, the aggregation of
starving slime mold amoebae \cite{KJLee96}, and the
Belousov-Zhabotinsky chemical reaction \cite{Belmonte96}.  Many of these
systems can be classified as excitable media in which the core of the
spiral, like a pacemaker, selects the temporal and spatial evolution
of the outward travelling spiral waves \cite{Belmonte96}.  In this
Letter we present a detailed experimental study of a driven, dissipative
system in which the formation of spirals can be attributed to a
qualitatively different mechanism acting far away from the spiral's
core. In particular, we show for Rayleigh-B\'{e}nard convection (RBC)
of a small Prandtl number fluid that the rotation of giant,
multi-armed spirals can be captured using concepts based on
nonlinear phase equations \cite{T86,NPS90,CT95,C97,LXG96,T98}.
These concepts should be universal and preliminary evidence
indicates that similar reasoning may also apply to the spiral pattern
formation in vibrating granular layers \cite{DBSGSS98} and in gas
discharges \cite{AMAP98}.

As stated earlier \cite{BDAC91}, the rigid rotation of a giant, finite
spiral of radius $r_d$ necessitates that the spiral waves which
propagate from the spiral's core are annihilated at $r = r_d$ by a
circular motion of the outer defect.  Thus, the pattern simply
consists of stationary, concentric rolls for $r > r_d$.  This
balancing mechanism has been placed on a more precise theoretical framework
by Cross and Tu \cite{CT95,C97} (CTC).  They argue that the rotation
of a spiral requires the reconciliation of two competing
selection principles acting far away from the spiral's core:
(1) wavelength selection by climb  of the outer defect
and (2) the emission of radially travelling waves due to target
selection.  These arguments have been successfully tested within
generalized Swift-Hohenberg (SH) models for axisymmetric spirals
\cite{CT95,C97,LXG96,T98} and have also been extended to multi-armed
spirals \cite{LXG96}.  However, the generalized SH-model used in these
studies contains ad~hoc 
parameters, and its stability regime deviates substantially from
that for RBC \cite{C74}.  As Cross \cite{C97} has pointed out in his
concluding remarks, serious conceptual uncertainties remain as well.

In this Letter, we present the first quantitative experimental and
theoretical analysis of giant rotating spirals in RBC. For
axisymmetric, multi-armed spirals in large aspect-ratio cells we find
good quantitative agreement between experimental measurements and
theoretical predictions.
However, for the frequent case of non-axisymmetric spirals our
measurements are in conflict with the proposed target selection
mechanism and, as a consequence, CTC's ``Invasive Chaos''
idea as a tentative explanation of Spiral Defect Chaos
(SDC) \cite{MBCA93} requires refinement.

\begin{figure}
\centerline{\epsfxsize=2.5in \epsfbox{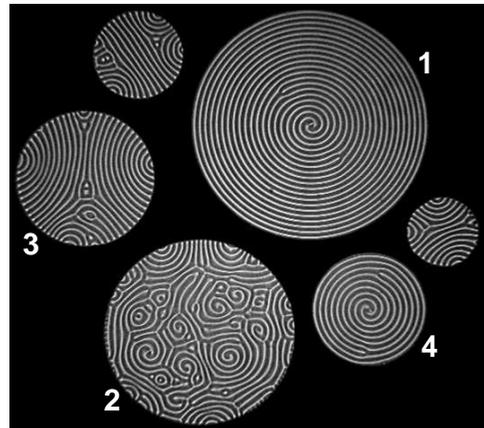}}
\caption{Shadowgraph image of the six convection cells for $\epsilon = 0.98$.
  Dark corresponds to warm up-flow, while light corresponds to
  cold down-flow. The
  cells used in the analysis are numbered from 1 to 4.  Cell 3 contains
  a PanAm pattern and cell 2 contains Spiral Defect Chaos.}
 \label{cells-fig}
\end{figure}

Rayleigh-B\'{e}nard convection occurs in a horizontal fluid layer of
height~$d$ heated from below when the temperature difference~$\Delta
T$ exceeds a critical value~$\Delta T_c$. For $\epsilon = (\Delta T /
\Delta T_c - 1) > 0$, a pattern of convection rolls with wavenumber $k
\approx \pi/d$ develops \cite{C74}.  Recent large aspect-ratio
experiments using pressurized gases revealed interesting
nonrelaxational pattern evolution.  For moderate $\epsilon$
Bodenschatz et al. \cite{BDAC91} found giant rotating spirals  similar
to those described here, while Morris et al.  \cite{MBCA93} found the
spatiotemporal chaotic state of spiral defect chaos (SDC).

As shown in \figref{cells-fig}, our experiment consisted of six
circular convection cells filled with pressurized ${\rm CO}_2$
gas. The convective pattern was visualized with the shadowgraph
technique \cite{deBruyn96}.  The experimental setup including 
parameters is described in
detail in Ref. \cite{PB96}. In most experimental runs reported here we
increased $\epsilon$ quasi-statically starting from below onset of
convection ($\epsilon = 0$).
Upon crossing onset, a small amplitude
axisymmetric target patterns developed in all cells due to weak static
sidewall forcing. (We achieved this by a step in the cells' sidewalls
\cite{PB96}).  Above $\epsilon \approx 0.4$ the initially axisymmetric
target moved off-center, compressing the pattern on one side
while dilating it on the other.  The appearance of this instability is
consistent with theoretical predictions
\cite{NPS90}. Upon further increases in $\epsilon$ the target
moved farther off-center until the wavenumber in the compressed region
increased beyond the skewed-varicose (SV)
instability \cite{C74} and a defect pair  would nucleate to
decrease the wavenumber.
One of the defects then moved to the center while the other
glided radially outward before coming to rest at a distance $r_d$ from
the geometrical center.  The pattern then relaxed to an on-center,
one-armed rotating spiral of radius $r_d$.  (While this
instability was observed for $\epsilon \approx 0.5$ in the largest
cell, it was slightly postponed in the smaller cells).
Above $\epsilon \approx 0.55$ we observed a behavior reminiscent of the
target instability in which the spiral's core would move off-center.
For $\epsilon < 0.64$ we observed stable, rotating, off-center
spirals (see Fig. 3
below) which with each increase in $\epsilon$ would
move further off-center. Eventually,  the wavenumber in the
compressed region increased beyond the SV-instability and defect
pairs nucleated. The pattern then developed
into a three- or four-armed
spiral, into the so-called PanAm pattern, or into Spiral Defect
Chaos. Examples of the latter two are shown in \figref{cells-fig} in cell
3 and cell 2, respectively.  We note that stability properties of spirals
had not been addressed experimentally nor theoretically prior to our
investigations.

For a few additional runs, we jumped the control
parameter from below
the onset of convection ($\epsilon < 0$) to above ($\epsilon > 0$).
As the jump was increased, we observed targets, one-armed spirals,
multi-armed spirals, PanAm patterns, and SDC.
Interestingly, we observed two-armed spirals only when we employed
this procedure.  Otherwise, the general trend with $\epsilon$ observed using
the two methods was similar.
We note that when $\epsilon$ was quasi-statically decreased starting
from SDC or PanAm patterns, we observed PanAm patterns, not targets,
even close to onset.  This shows that the static sidewall
forcing was very weak.

\begin{figure}
\centerline{\epsfxsize=2.4 in\epsfbox{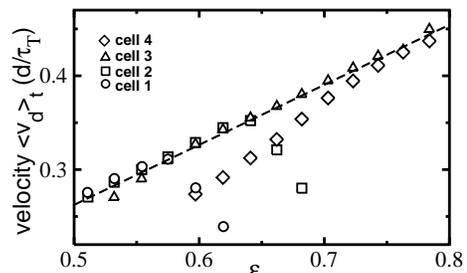}}
\caption{Average velocity $v_d$ of the outer defect {\em vs.}~$\epsilon$ for
one-armed spirals.  The size of the on-center spirals is given in Ref.
\protect\cite{radius}. The dashed line  represents a linear fit to the
on-center data.  }
\label{velocity}
\end{figure}

Let us initially focus on the
one-armed spiral.  \figref{velocity} shows the average velocities
$v_d$ of the outer defects for an
experimental run with one-armed spirals in cells 1--4.  (The velocities are
normalized in terms of $d/\tau_T$, where $\tau_T$ is the vertical thermal
diffusion timescale.)
Note that the
spirals had a variety of different sizes $r_d$.  We measured $v_d$
by tracking the path of the outer defect
with a Fourier demodulation technique \cite{GPRS89,P97}. 
For on-center spirals the averaged
defect velocities of spirals in cells 1--3 obeyed the same linear
relationship $v_d = 0.64 (\epsilon-0.09)$ .
For values of $\epsilon$ with off-center spirals,
$v_d$ changed abruptly.  (The spiral in cell 4
showed deviations from this behavior, possibly due to its small size.)
We also measured the rotation frequency $\omega$ of
the spirals and found that the data are well-described by $\omega = v_d
/ r_d$. This dependence on spiral size differs markedly from the
size-independence of $\omega$ observed for spirals in excitable media
\cite{Belmonte96}.

The first part of CTC's argument is that the behavior of the outer
defect of a spiral can be considered as a dislocation climbing in a
(slightly curved) roll pattern with a radial wavenumber $q(r)$ in
analogy to the simpler case of a dislocation climbing with the
velocity $v_d$ in a straight roll pattern with background wavenumber
$q$.  This latter case was considered theoretically in the framework
of the Swift-Hohenberg model \cite{T86}.  Based on the phase diffusion
equation which captures the behavior far from the defect, the defect
velocity $v_d$ was found to obey the relation \cite{Croq-comment}:
\begin{equation}
v_d = \beta (\epsilon) (q-q_d(\epsilon)),
\label{vdbetaqqd}
\end{equation}
where $q_d$ is the zero-velocity wavenumber.

In order to compare with this theoretical prediction one first needs a
sensible definition of a background wavenumber $q(r)$.
Crucial to this was the observation that the time-average of an
$m$-armed spiral (when performed over a duration equal to a multiple of
the rotation period)  yielded a target pattern.
This is
exemplified in \figref{average}(B) for the one-armed off-center spiral
shown in \figref{average}(A).  As shown below,  choosing  the radial
wavenumber $q(r)$ of the target pattern as
the background wavenumber $q$  in \eqref{vdbetaqqd} appeared
to work quite well.

The occurrence of a target pattern after averaging over a rotating
spiral can be rationalized by the following approximate
calculation. A one-armed, finite spiral can be described satisfactorily by a
modified Archimedean spiral, {\it i.e.}, $f(\vec{r},t) = A(r) \cos
(q\,r - \phi + \phi_d - \omega t)$, where $q$ is the wavenumber of the
underlying target without defects, $A(r)$ is the amplitude, and $r$ is
the radial distance from the spiral's center. The phases $\phi =
\arctan(y/x)$ and $\phi_d = \arctan((y-y_d)/(x-x_d) $ are polar angles
centered about the spiral's core at $\vec{r}=(0,0)$ and the outer
defect position $\vec{ r}_d(t)=(x_d,y_d)$, respectively.  It is not
difficult to show that the time-average of such a rigidly rotating
spiral gives $ I(r,r_d) = 2 \pi A(r) J_1\left(r/r_d\right) \cos(q r)$,
where $J_1$ is the first Bessel function of the first kind.
We note that the corresponding average of an Archimedean
spiral ($\phi_d =0 $) would
vanish.

\begin{figure}
\centerline{\epsfxsize=2.7in \epsfbox{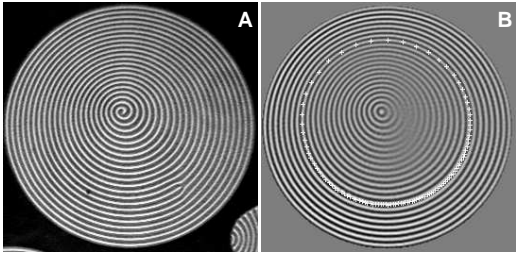}}
\caption{(A) Off-center one-armed spiral and (B) the average over one
rotation period overlaid with the defect positions at
constant time intervals $\Delta t = 8.7 \tau_T$ for $\epsilon = 0.62$
and $\sigma = 1.38$.
}
\label{average}
\end{figure}

\begin{figure}
\centerline{\epsfxsize=2.6in\epsfbox{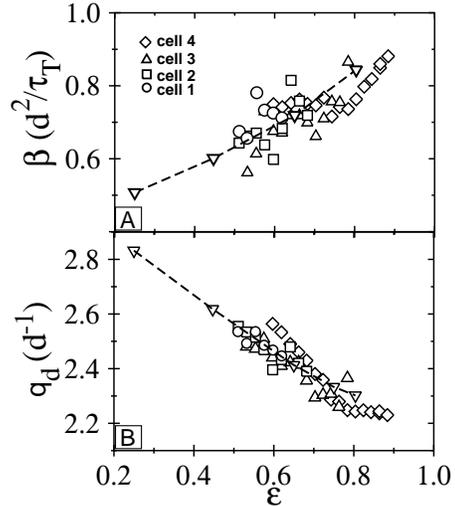}}
\caption{ (A) $\beta$ and (B) $q_d$ {\em vs.}~$\epsilon$ for single-armed,
 off-center spirals, compared with
 numerical simulations of defects in a straight roll pattern. }
\label{beta_qd}
\end{figure}

Using the above definition of the background wavenumber $q(r)$,
we tested \eqref{vdbetaqqd} using experimental
velocities obtained from off-center, one-armed spirals
over a wide range of values of $\epsilon$.
The slower defect motion in the dilated regions and the faster motion
in the compressed regions seen in Fig. 3B is consistent with
\eqref{vdbetaqqd}. For a quantitative test we measured $q(r)$ at the defect
positions by fitting small regions to concentric
roll patterns \cite{P97}.  (The method of Ref. \cite{Egolf98} produced
similar values.)  We then plotted $v_d$ versus $q$ for each value of
$\epsilon$ and found that the data were well-described by a linear
relationship, allowing us to determine the parameters
$\beta(\epsilon)$ and $q_d(\epsilon)$ using a least-squares fitting
procedure \cite{extracomment}.
These data are shown in \figref{beta_qd}.  Using
simulations of the three-dimensional Boussinesq equations \cite{P96},
we also measured defect velocities $v_d$ as a function of background
wavenumber $q$ for defects moving in straight roll patterns.
Again, we found a linear relationship between $v_d$ and $q$ and
determined $\beta(\epsilon)$ and $q_d(\epsilon)$.  As shown in
\figref{beta_qd}, the numerically determined values for
$\beta(\epsilon)$ and $q_d(\epsilon)$ are in excellent agreement with
the experiment.  Using the simulations, we also analyzed off-center,
one-armed spirals \cite{Dressel} and found excellent agreement between
experiment and theory \cite{log-comment}.

The second part of CTC's argument --- selection by circular traveling
waves --- relies on the fact that away from the core of the spiral the
wavefronts deviate only slightly from circular and are thus
well-approximated as targets.  It has been shown that targets prefer a
specific wavenumber $q_t(\epsilon)$ \cite{NPS90}, and that a target
with a wavenumber differing from $q_t(\epsilon)$ will attempt to
adjust its wavenumber by emitting circular waves of frequency
$\omega_t(\epsilon)$.  Using the nonlinear Cross-Newell
phase-diffusion equation \cite{NPS90}, one finds
\begin{equation}
    \omega_t   =   \alpha(\epsilon) (q_t(\epsilon) - q)/r,
\label{alpha}
\end{equation}
where $\alpha = 2D_\parallel(q_t)$, $D_\parallel(q)$ is the parallel
diffusion constant, and $r$ is the distance from the center of the
target. The numerical value of the parameter $\alpha$ can be
calculated from the growth rate $\sigma(q_t, K)$ of a longitudinal
modulation with wavenumber $K$ of a pattern with wavenumber
$q_t$ as $\alpha = (-2 \sigma(q_t, K) /
{K^2}){\mid_{K\rightarrow 0}}$ \cite{NPS90}.

\begin{figure}
\centerline{\epsfxsize=2.7in \epsfbox{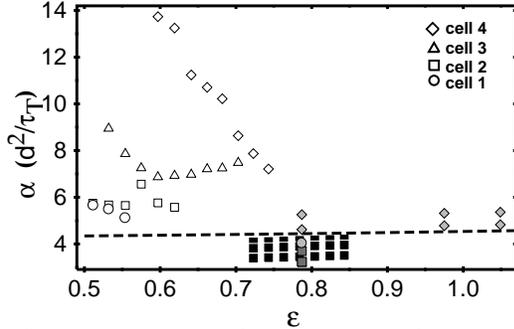}} 
\caption{
$\alpha$ {\em vs.}~$\epsilon$
for single- and multi-armed spirals.
The shapes of the symbols correspond to the four different cells.  The
level of gray in the symbols distinguishes between the number of arms
in the spirals.  One-armed spirals are clear, two-armed are gray,
three-armed are dark gray, and four-armed are black. The numerically
determined value of $\alpha$ is plotted as a dashed line. }
\label{alpha1} \end{figure}

Using geometrical arguments, CTC showed that
a one-armed spiral requires:
\begin{equation}
     v_d =  \omega_t  r_d,
\label{mv}
\end{equation}
where, again, $v_d$ is the velocity of the defect at radius $r_d$.  We
used \eqref{mv} and \eqref{alpha} and the generalization to multi-armed
spirals \cite{LXG96} to determine $\alpha$.  From the investigations
above and a similar analysis for $m$-armed, on-center spirals, we
measured the average defect velocity $v_d$. For each cell we
determined $q_t$ by extrapolating the azimuthal average of $q_t
(\epsilon)$ at the defect positions from on-center targets at lower
$\epsilon$ to the larger $\epsilon$ where we observed the $m$-armed,
on-center spirals.  We found that $(q_t-q_c) \sim \epsilon$ \cite{P97}. The
data for $q_t$ was in good agreement with the numerical predictions
by Buell and Catton \cite{Buell}.  The
remaining unknown background wavenumber $q$ was measured from the
underlying target, again azimuthally averaged over the defect
positions.  With this information we determined $\alpha$ for one-, two-,
three-, and four-armed on-center spirals. The data are summarized in
\figref{alpha1}.  It shows that for the larger cells and the
multi-armed spirals $\alpha$ agrees well with the theoretical value
numerically determined  from the growth rate of the Eckhaus instability
using a Galerkin code (dashed
line). The discrepancies become significant for one-armed spirals in
the smaller cells 3 and 4.

For the frequently occurring off-center spirals we found that the
rotating spirals had regions where the local wavenumber was larger
than that of the stationary target.  This observation indicates that a
``local'' interpretation of Eq. (2) (negative $\omega_t$) is not
reasonable. Rotating spirals are important structure elements of SDC
\cite{MBCA93,Egolf98}.  Mostly they are one-armed and often strongly
off-center.  In this case the target selection becomes unreliable and
as a consequence CTC's ``Invasive Chaos'' idea as an explanation of
SDC has to be reconsidered.

In summary, our analysis presents the first quantitative test of
\eqref{vdbetaqqd} \cite{T86} and  \eqref{alpha} \cite{CT95,C97,LXG96}
in a real physical system.
Giant multi-armed spirals are exceptionally well-suited for
the study of dislocation dynamics in RBC; one can follow the defect
trajectories for very long times and, in addition, the background
wavenumber is naturally defined by the time average of the spiral.

This work was supported by the National Science Foundation
(DMR-9320124, ASC-9503963 ,and DMR-9705410), by NATO (CRG-950243), and
by the Cornell Theory Center.  B.B.P acknowledges support from the
Department of Education and W.P. thanks the Cornell Matrial Science
Center for its hospitality.





\begin{thebibliography}{10}
\bibitem{radius} The radius of the outer defect was for cell 1: $r_d =
\{(34.1 - 2.0 \epsilon) \pm0.3\}d$ for $0.51 < \epsilon < 0.55$ ; cell
2: $r_d = \{(30.7 - 1.5 \epsilon) \pm0.3\}d$ for $0.51 < \epsilon <
0.64$; cell 3: $r_d = \{(20.0 - 1.2 \epsilon) \pm0.3\}d$ for $0.53 <
\epsilon < 0.78$; cell 4: $r_d = \{(14.8 - 0.7 \epsilon) \pm0.3\}d$
for $0.60 < \epsilon < 0.76$.
\bibitem{Croq-comment} \eqref{vdbetaqqd} is valid for a {\sl non-potential}
system.  Pocheau and Croquette
\cite{Pocheau} measured the defect dynamics in an experimental straight roll
pattern for $\sigma = 70$ and found that their results agreed with the
prediction for a potential system where $v_d \sim (q-q_d)^{2/3}$ (see
\cite {T86}).
\bibitem{extracomment} The linear fit
was appropriate for the most  off-center
spirals, but for  spirals whose variation in wavenumber was
small,  the linear portion of the $v_d$ versus $q$ plot was
chosen for the fit (for details see
Ref.~\protect\cite{P97}).
\bibitem{log-comment} We note that a weak logarithmic correction
to \eqref{vdbetaqqd} was predicted by Tusaro and Cross \cite{T86}; however, 
our data did not allow us to verify this small correction.
\end{thebibliography}
\end{document}